\documentclass[twocolumn,aps,prl,showpacs,preprintnumbers]{revtex4}%
\usepackage{mathrsfs}
\usepackage{amsmath}
\usepackage{amsfonts}
\usepackage{amssymb}
\usepackage{amsthm}
\usepackage{graphicx}
\usepackage{color}
\usepackage{txfonts}%
\providecommand{\U}[1]{\protect\rule{.1in}{.1in}}
\begin{document}
\title{Angular and Linear Momentum of Excited Ferromagnets}
\author{Peng Yan$^{1}$}
\author{Akashdeep Kamra$^{1}$}
\author{Yunshan Cao$^{1}$}
\author{Gerrit E.W. Bauer$^{2,1}$}
\affiliation{$^{1}$Kavli Institute of NanoScience, Delft University of Technology,
Lorentzweg 1, 2628 CJ Delft, The Netherlands}
\affiliation{$^{2}$Institute for Materials Research and WPI-AIMR, Tohoku University, Sendai
980-8577, Japan}

\begin{abstract}
The angular momentum\ vector of a Heisenberg ferromagnet with isotropic
exchange interaction is conserved, while under uniaxial crystalline anisotropy
the projection of the total spin along the easy axis is a constant of motion.
Using Noether's theorem, we prove that these conservation laws persist in the
presence of dipole-dipole interactions. However, spin and orbital angular
momentum are not conserved separately. We also define the linear momentum of
ferromagnetic textures. We illustrate the general principles with special
reference to the spin transfer torques and identify the emergence of a
non-adiabatic effective field acting on domain walls in ferromagnetic insulators.

\end{abstract}

\maketitle

Mathematics can be very effective in guiding research when physical intuition
fails, even in applied sciences such as condensed matter physics. An important
tool is Noether's theorem \cite{Noether} that helps identifying invariants or
continuity equations from the fundamental symmetry properties of a given
system. In the field of spintronics, for instance, Noether's theorem has been
used to express the spin current, i.e. the flow of spin angular momentum
\cite{Maekawa}, in spin-orbit-coupled systems \cite{Xia}. In metallic
ferromagnets spin current is carried by an imbalance between up-spin and
down-spin electrons and therefore accompanied long-distance mass motion and
strong Joule heating. Spin currents can also be carried by spin waves
(magnons), thereby dissipating much less energy in some magnetic insulators
with high crystal quality \cite{Maekawa}. Magnon mediated spin-current
transport in various systems has received some attention in the past years
\cite{Takigawa,Zotos,Alvarez,Loss}. Sch\"{u}tz \emph{et al.} \cite{Schutz1}
demonstrated that magnons in a mesoscopic Heisenberg ring generate a
persistent spin current under an inhomogeneous magnetic field. In
magnetization textures particle-based \cite{Parkin,Tatara,Zhang} as well as
magnonic spin currents \cite{Mikhailov,Yan,Hinzke,Kovalev} cause spin transfer
torques that induce magnetization dynamics such as a domain wall (DW) motion.
Direct imaging of a domain wall motion induced by thermally induced magnonic
spin currents has been reported by Jiang \emph{et al} \cite{Jiang}. The spin
transfer torque in magnetic insulators is usually ascribed to conservation of
spin angular momentum, implicitly assuming that the exchange interaction is
isotropic. However, whereas a negative domain wall velocity, i.e. opposite to
the spin wave propagation direction, is the signature of a magnonic spin
transfer torque \cite{Yan,Hinzke,Kovalev}, positive domain wall velocities
were found in micromagnetic simulations \cite{Han,Jamali,Seo,Kim,Wang}. A
conclusive explanation of the latter observation is still lacking. Even the
spin current and the corresponding continuity equation in Heisenberg magnets
has not yet been properly formulated \cite{Schutz2}.

Tatara and Kohno \cite{Tatara} predicted domain wall motion by the force or
(linear) momentum transfer felt by narrow domain walls at which electron spins
are reflected. But Volovik \cite{Volovik} noted that the linear momentum of
magnetization dynamics is not invariant under spin rotations \cite{Haldane}
and explained this paradox by considering a dynamic equation for the spin
degrees of freedom supplemented by a kinetic equation for the underlying
incoherent fermionic excitations \cite{Volovik,Yaroslav}. However, this
approach fails for ferromagnetic insulators, illustrating the need for a full
understanding of linear and angular momentum transport in ferromagnets.

In this Letter, we formulate the angular and linear momentum of excited
ferromagnets based on Noether's theorem \cite{Noether}. Starting with the
Landau-Lifshitz equation for magnetization dynamics and Maxwell's equations
for dipolar fields, we provide a systematic formulation of the conservation
laws for the rotational and translational motion of spin excitations, e.g.,
magnons or domain walls, based on general symmetry principles. We show that in
the presence of magnetic dipole-dipole interactions the spin current is not
conserved, only the total angular momentum composed of spin and orbital
component is. Noether's theorem also leads us to a proper formulation of
linear momentum in ferromagnets that identifies the non-dissipative linear
momentum transfer (\textquotedblleft effective field\textquotedblright)
mechanism in magnetic textures.

The semiclassical dynamics of a ferromagnet is described by the
Landau-Lifshitz equation\textit{ }
\begin{equation}
\frac{\partial\mathbf{M}}{\partial t}=-\mathbf{M}\times\mathbf{H}_{eff},
\label{Landau}%
\end{equation}
where $\mathbf{M}=\left(  M_{x},M_{y},M_{z}\right)  $ is the magnetization
vector with modulus $M_{0}=\left\vert \mathbf{M}\right\vert $ and
$\mathbf{H}_{eff}=-\delta E/\delta\mathbf{M}$ is the effective field expressed
as the variational derivative of the energy $E=\int\mathcal{H}dV$. The energy
density
\begin{equation}
\mathcal{H=}\frac{J}{2}\left(  \nabla\mathbf{M}\right)  ^{2}+f\left(
M_{z}\right)  -\mathbf{M\cdot h}-\frac{\mathbf{h}^{2}}{8\pi}%
\end{equation}
consists of the exchange interaction, the magnetic dipole interaction
expressed by the field $\mathbf{h}$, and we chose here an easy uniaxial
anisotropy $f$ along the $z$-axis. $\mathbf{h}$ obeys Maxwell's equations but
for slow modulations considered here the magnetostatic approximation suffices
\cite{Stancil}:
\begin{equation}
\nabla\times\mathbf{h}=0;\;\nabla\cdot\left(  \mathbf{h}+4\pi\mathbf{M}%
\right)  =0. \label{Maxwell}%
\end{equation}
We can write the Lagrangian density of the system as%
\begin{equation}
\mathcal{L}=-M_{z}\dot{\phi}-\mathcal{H}, \label{Lagrangian}%
\end{equation}
where $\phi=\arctan\left(  M_{y}/M_{x}\right)  $ is the azimuthal angle of
$\mathbf{M}$. With $\mathbf{h}=\nabla\psi,$ the first of Eqs. \eqref{Maxwell}
is satisfied identically, while Eq. \eqref{Landau} and the second equation of
\eqref{Maxwell} are the Euler-Lagrange equations%
\begin{equation}
\frac{\partial}{\partial x_{i}}\frac{\partial\mathcal{L}}{\partial\left(
\partial q/\partial x_{i}\right)  }=\frac{\partial\mathcal{L}}{\partial q},
\label{EulerLagrange}%
\end{equation}
where $q=M_{z},\phi,\psi;$ $i=\left\{  1,\ldots,4\right\}  ;$ $x_{1,2,3}%
=x,y,z;$ $x_{4}=t.$

We can now construct field invariants, i.e., a combination of the fields and
their derivatives as functions of time and space that is conserved in time
\cite{Bogolyubov}. According to Noether's theorem any continuous
transformation of coordinates under which the variation of the action vanishes
generates a definite invariant. We employ the global symmetries to obtain
conservation laws for a closed system containing a magnetization texture and
the associated dipolar field.

\textit{Translational symmetry.}$\ $Spatial translational invariance leads to
the conservation of linear momentum while a time translation symmetry gives
rise to energy conservation. Application of Noether's theorem leads to the
continuity equation $\partial T_{ik}/\partial x_{k}=0$ for the energy-momentum
tensor
\begin{equation}
T_{ik}=\left(  \frac{\partial q}{\partial x_{i}}\frac{\partial}{\partial
\left(  \partial q/\partial x_{k}\right)  }-\delta_{ik}\right)  \mathcal{L},
\label{EMTensor}%
\end{equation}
which can be derived from the invariance of the action under the
spatiotemporal translation transformations $\delta x_{i}=\delta_{ij}%
\delta\epsilon_{j}$ and $\delta M_{z}=\delta\phi=\delta\psi=0,$ where
$\delta_{ij}$ is the Kronecker function and $\delta\epsilon_{j}$ are
infinitesimal translations. $T_{44}$ is the energy density and $T_{i4}$ the
linear momentum density%
\begin{equation}
p_{i}=T_{i4}=-M_{z}\frac{\partial\phi}{\partial x_{i}}. \label{LMD}%
\end{equation}
Hence, the total energy $E=\int T_{44}dV$ and linear momentum%
\begin{equation}
P_{i}=\int T_{i4}dV \label{TL}%
\end{equation}
are conserved.

This conservation law is complicated by the non-differentiability of the
azimuthal angle $\phi$ at the north $\left(  \theta=0\right)  $ or south poles
$\left(  \theta=\pi\right)  $ \cite{Papanicolaou}. By parameterizing the spin
variables in terms of $\left(  M_{x},M_{y},M_{z}\right)  \ $the momentum
density $\left(  1-\cos\theta\right)  \frac{\partial\phi}{\partial x_{i}}$ can
be written as $\mathbf{A}\cdot\frac{\partial\mathbf{M}}{\partial x_{i}},$
where $\mathbf{A}=\left(  M_{x}\mathbf{e}_{y}-M_{y}\mathbf{e}_{x}\right)
/M_{0}\left(  M_{0}+M_{z}\right)  \ $diverges on the line specified by the
equations $M_{x}=M_{y}=0$ and $M_{z}=-M_{0}$ (south pole). The singularity can
be removed by employing an arbitrariness in the Lagrangian \eqref{Lagrangian}
that can be written as $\mathcal{L}=\left(  C-M_{z}\right)  \dot{\phi
}-\mathcal{H}$, where the choice of the constant $C$ does not affect the
dynamics. While a given $C$ cannot remove the singularities at two poles
simultaneously. \begin{figure}[ptbh]
\begin{center}
\includegraphics[width=8.5cm]{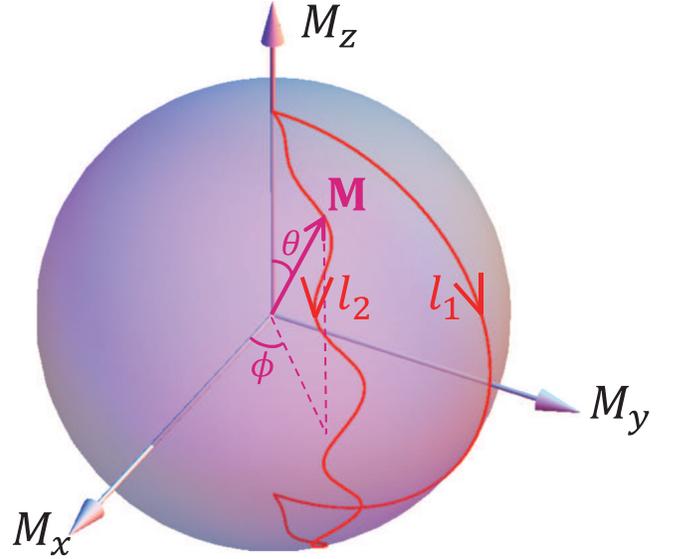}
\end{center}
\caption{(Color online) The Bloch sphere $\left\vert \mathbf{M}\right\vert
\ =M_{0}$ with trajectories through domain walls with various configurations.
Trajectory $l_{1}$ indicates a planar domain wall in the $y-z$ plane, while
$l_{2}$ describes a general domain wall structure with local twists. The area
of the contour $l_{2}\bar{l}_{1}$ on the sphere specifies the difference in
the domain wall momentum described by trajectories $l_{2}$ and $l_{1}$. }%
\label{Firstfigure}%
\end{figure}

The dynamic part of Lagrangian \eqref{Lagrangian} in terms of $\mathbf{M}$ and
$\partial\mathbf{M}/\partial t$ coincides with that of a charged particle in a
non-singular magnetic field $\mathbf{B}=\nabla\times\mathbf{A}=-\mathbf{M}%
/M_{0}^{3}$ for $M_{0}\neq0,$ in terms of a vector potential $\mathbf{A}$.
Varying $C$ is therefore equivalent to a gauge transformation. The linear
momentum $p_{i}=\left(  C-M_{z}\right)  \partial\phi/\partial x_{i}$ is not
invariant under these gauge transformations, but the difference between
momenta of distinguishable states is. We illustrate this notion by the
momentum of a $180^{\circ}$ magnetic domain wall\ that is determined by path
integrals of the form $M_{0}\int\mathbf{A}\cdot d\mathbf{M}$ along a
trajectory connecting $M_{z}=-M_{0}$ and $M_{z}=+M_{0}$ [see $l_{1}$ or
$l_{2}$ in Fig. \ref{Firstfigure}]. The difference between the momenta is
governed by the integral $M_{0}\int\mathbf{A}\cdot d\mathbf{M}$ along a closed
contour, i.e., path $l_{2}\bar{l}_{1}$. According to Stoke's theorem, the
integral in question can be represented as the flux of the vector
$\mathbf{B}=\nabla\times\mathbf{A}$ through the surface bounded by this
contour%
\begin{equation}
\Delta P_{2-1}=M_{0}\iint_{S_{2-1}}\mathbf{B}\cdot d\mathbf{S},
\label{MomentumDifference}%
\end{equation}
which is now gauge invariant. The spin electromotive force for electrons in
moving magnetization textures is expressed by a similar contour integral
\cite{Barnes}. Here we express the DW momentum as
\begin{equation}
P_{\mathrm{DW}}=M_{0}\iint\sin\theta d\theta d\phi\label{DWmomentum}%
\end{equation}
where the momentum of a fully in-plane DW $\left(  \phi=0\right)  $ defines
the zero. When the DW plane is not twisted, the above equation leads to
$P_{\mathrm{DW}}=2\phi M_{0}.$ The linear momentum carried by a rigid DW only
depends on the tilt angle $\phi$ of its plane. The conclusion that even a
completely static domain wall has a finite linear momentum is
counter-intuitive but can be rationalized in terms of the persistent angular
momentum current generated by this topological defect.

Let us now consider the reflection of a spin wave with wave vector
$\mathbf{k}$ by a planar domain wall as illustrated in Fig. 2(a). According to
Eq. \eqref{TL}, the total linear momentum should be conserved during the
scattering process%
\begin{equation}
0=\frac{d\mathbf{P}}{dt}=\frac{d}{dt}\left(  \mathbf{P}_{\mathrm{SW}%
}+\mathbf{P}_{\mathrm{DW}}\right)  .
\end{equation}
where $\mathbf{P}_{\mathrm{SW}}$ and $\mathbf{P}_{\mathrm{DW}}$ are the
momenta of spin wave and domain wall, respectively. Therefore, we have%
\begin{equation}
\mathbf{F}_{\mathrm{DW}}=2M_{0}\dot{\phi}\mathbf{e}_{z},\label{Force}%
\end{equation}
where $\mathbf{F}_{\mathrm{DW}}=d\mathbf{P}_{\mathrm{DW}}/dt$ is the force
transferred by the spin wave to the domain wall by reflection. We find that
spin wave reflection is possible only under simultaneous rotation of the
domain wall. \textit{Vice versa}, linear momentum transfer does not lead to
linear motion of the domain wall, but a rotation of the domain wall plane.
Linear momentum transfer is thereby shown to be equivalent to an
\textquotedblleft effective\textquotedblright\ Zeeman magnetic field. Note
that when the axial symmetry is broken, spin wave reflection without coherent
rotation becomes possible. \begin{figure}[ptbh]
\begin{center}
\includegraphics[width=8.5cm]{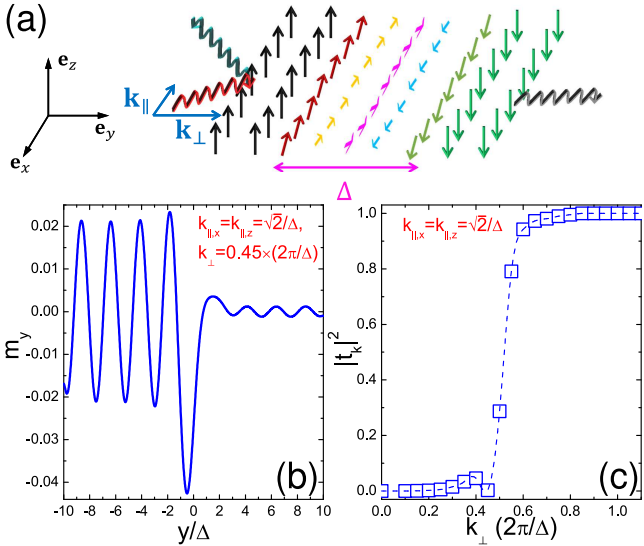}
\end{center}
\caption{(Color online) (a) Illustration of spin waves (wavy lines with
arrows, red for incoming, aqua for reflection, and gray for transmission)
scattered by a Bloch wall. $\mathbf{k}_{\parallel}$ and $\mathbf{k}_{\perp}$
stand for the wave vectors parallel and perpendicular to the domain wall
plane, respectively. $\Delta$ is the domain wall width. (b) Spin wave
reflection by a DW in 3D with $\mathbf{k}_{\parallel}=\left(  \sqrt{2}%
/\Delta\right)  \left(  \mathbf{e}_{x}+\mathbf{e}_{z}\right)  $ and
$\mathbf{k}_{\perp}=0.45\times\left(  2\pi/\Delta\right)  \mathbf{e}_{y}$
obtained by solving the linearized Landau-Lifshitz equation coupled with
Maxwell's equation \cite{Prepare}. The anisotropy energy is $f\left(
M_{z}\right)  =-\frac{1}{2}KM_{z}^{2}$ with $K/4\pi=0.01.$ (c) $k_{\perp}$
dependence of SW transmission $\left\vert t_{k}\right\vert ^{2}$ for a fixed
$\mathbf{k}_{\parallel}$.}%
\end{figure}

When including energy dissipation in the form of Gilbert damping
$\alpha\mathbf{M}\times\mathbf{\dot{M}/}M_{0}$ by a collective coordinate
approach and Walker ansatz \cite{Walker}, we find that rotation of the DW
plane induced by spin wave reflection is associated with linear propagation
along the same direction as the spin wave. Its velocity $\mathbf{v}%
_{\mathrm{DW}}=\alpha\Delta\dot{\phi}\mathbf{e}_{z}$, where $\Delta$ is the DW
width. We thereby solve the puzzles mentioned in the introduction
\cite{Han,Jamali,Seo,Kim,Wang}.\ Spin wave reflection at a domain wall is per
definition a \textquotedblleft non-adiabatic\textquotedblright\ process. The
resultant torque due to this linear momentum transfer is non-dissipative,
thereby different from the dissipative one proposed by Kovalev and Tserkovnyak
\cite{Kovalev}. Unlike the dissipative correction for the adiabatic limit of
wide domain walls \cite{Kovalev}, the non-dissipative one predicted here is
significant for domain walls subject to spin waves with non-normal incidence
$\left(  k_{\parallel}\neq0\right)  $ and $k_{\bot}\Delta\lesssim0.1$. The
latter statement requires some qualification. Spin wave reflection predicted
in a 1D spin chain \cite{Peng} (equivalent to a normal incidence spin wave in
higher dimensions) requires atomic pinning by the discrete lattice, an effect
that is beyond the continuum model used here. However, for $k_{\parallel}%
\neq0$ and sharp domain walls the dipolar interaction leads to strong
reflection also in the continuum model as is illustrated by the computed spin
wave amplitudes in Fig. 2(b), while the $k_{\perp}$ dependence of spin wave
transmission probabilities is presented in Fig. 2(c) \cite{Prepare}.\textit{
}The dissipative correction favours a negative DW velocity (opposite to
$\mathbf{k}_{\bot}$) \cite{Kovalev}, while the non-dissipative non-adiabatic
torque leads to a positive one. The two\ mechanisms can be distinguished in a
series of experiments or simulations on magnetic wires in which either the
domain wall widths or the wavelengths of injected magnons is tunable.

\textit{Rotational symmetry.}$\ $According to Noether's theorem the axial
symmetry under spatial rotation around the easy $z$-axis, also in the presence
of magnetostatic dipole-dipole interaction, implies conservation of angular
momentum in this direction. A rotation around the $z$-axis is generated by
$\delta x=y\delta\epsilon,$ $\delta y=-x\delta\epsilon,$ $\delta z=\delta
M_{z}=\delta\psi=0,$ and $\delta\phi=-\delta\epsilon,$ where $\delta\epsilon$
is the infinitesimal rotation parameter around $z$-axis. Defining the
$z$-component of the angular momentum current density
\begin{align}
j_{z\mu}  &  =J\left(  M_{0}^{2}-M_{z}^{2}\right)  \frac{\partial\phi
}{\partial x_{\mu}}-J\left(  M_{0}^{2}-M_{z}^{2}\right)  \frac{\partial\phi
}{\partial x_{\mu}}\left(  \mathbf{r}\times\nabla\phi\right)  _{z}\nonumber\\
&  -J\frac{M_{0}^{2}}{M_{0}^{2}-M_{z}^{2}}\frac{\partial M_{z}}{\partial
x_{\mu}}\left(  \mathbf{r}\times\nabla M_{z}\right)  _{z}+M_{\mu}\left(
\mathbf{r}\times\nabla\psi\right)  _{z}\nonumber\\
&  +\frac{1}{4\pi}\frac{\partial\psi}{\partial x_{\mu}}\left(  \mathbf{r}%
\times\nabla\psi\right)  _{z}+\varepsilon_{\mu z\nu}x_{\nu}\mathcal{L,}
\label{Flux}%
\end{align}
where $\varepsilon_{\mu z\nu}$ is the Levi-Civita symbol and $\mu=\left\{
1,2,3\right\}  ,$ as well as
\begin{equation}
j_{z4}=M_{z}-M_{z}\left(  \mathbf{r}\times\nabla\phi\right)  _{z}, \label{AMD}%
\end{equation}
Noether's theorem leads us to the conservation law for the angular momentum
along the $z$-axis%
\begin{equation}
\frac{\partial j_{z4}}{\partial t}+\frac{\partial j_{z\mu}}{\partial x_{\mu}%
}=0, \label{AngularConservation}%
\end{equation}
The first term in Eq. \eqref{AMD} is the spin angular momentum density
\cite{Yan}, while the second one can be identified as an orbital angular
momentum density since it can be written as $\left(  \mathbf{r}\times
\mathbf{p}\right)  _{z}$ where $\mathbf{p}=-M_{z}\nabla\phi$ is the linear
momentum density obtained before [Eq. \eqref{LMD}]. Noether's theorem states
that $\int\left(  \partial j_{z\mu}/\partial x_{\mu}\right)  dV=0.$
Specifically, the $z$-component of the total angular momentum%
\begin{equation}
J_{z}=\int j_{z4}dV \label{TA}%
\end{equation}
is conserved.

Conservation of spin angular momentum has been discussed for purely
exchange-coupled ferromagnets \cite{Mikhailov,Yan}. In the presence of
magnetic dipolar interactions, the energy-momentum or stress tensor $T_{ik}$
becomes non-symmetric and the orbital angular momentum%
\begin{equation}
L_{i}=\int\varepsilon_{ijk}x_{j}T_{k4}dV,\text{ }i=\left\{  1,2,3\right\}  .
\end{equation}
does not vanish. The $z$-projection of the integrand, i.e. the orbital angular
momentum density, agrees with the second term of Eq \eqref{AMD}. Since
$dL_{i}/dt=-\int\varepsilon_{ijk}M_{j}\left(  \partial\psi/\partial
x_{k}\right)  dV\neq0,\ $a non-zero $L_{i}$ is then not conserved. We note the
analogy with the coupling of spins by the spin-orbit interaction of electrons
in the weakly relativistic limit, a role that is played here by the
dipole-dipole interaction. We note that the angular momentum density of the
electromagnetic field $\vec{j}=\mathbf{r}\times\left(  \mathbf{e}%
\times\mathbf{h}\right)  $ \cite{Allen} is negligibly small in the regime
where the magnetostatic approximation [as in Eq. \eqref{Maxwell}] holds, in
which the electric field $\mathbf{e}$ plays no role whatsoever \cite{Stancil}.

We now illustrate our results [Eq. \eqref{AMD}] for a uniform ferromagnetic
nanocylinder $\left(  \mathbf{M}=M_{0}\mathbf{e}_{z}\right)  $ with uniaxial
anisotropy $f\left(  M_{z}\right)  =-\frac{1}{2}KM_{z}^{2}$ and spin wave
excitation $\mathbf{m}=\left(  m_{x},m_{y},0\right)  $ $\left(  \mathbf{M}%
=M_{0}\mathbf{e}_{z}+\mathbf{m}\right)  $. The Hamiltonian to leading order in
$\mathbf{m}$ is diagonalized by the Bogoliubov transformation
\begin{align}
m_{+}  &  =m_{x}+im_{y}\nonumber\\
&  =\sqrt{\frac{2\hslash M_{0}}{V}}\sum_{\mathbf{k}}\left\{  u_{\mathbf{k}%
}a_{\mathbf{k}}e^{i\left(  \mathbf{k}\cdot\mathbf{r}-\omega_{\mathbf{k}%
}t\right)  }+\upsilon_{\mathbf{k}}^{\ast}a_{\mathbf{k}}^{+}e^{-i\left(
\mathbf{k}\cdot\mathbf{r}-\omega_{\mathbf{k}}t\right)  }\right\}  ,
\end{align}
where $a_{\mathbf{k}}^{+}$ and $a_{\mathbf{k}}$ are Bose creation and
annihilation operators while the coefficients $u_{\mathbf{k}}$ and
$\upsilon_{\mathbf{k}}$ and the frequency $\omega_{\mathbf{k}}$ are related by
the equations $A_{\mathbf{k}}u_{\mathbf{k}}+B_{\mathbf{k}}^{\ast}%
\upsilon_{\mathbf{k}}=\omega_{\mathbf{k}}u_{\mathbf{k}},$ $B_{\mathbf{k}%
}u_{\mathbf{k}}+A_{\mathbf{k}}\upsilon_{\mathbf{k}}=-\omega_{\mathbf{k}%
}\upsilon_{\mathbf{k}}$ and $\left\vert u_{\mathbf{k}}\right\vert
^{2}-\left\vert \upsilon_{\mathbf{k}}\right\vert ^{2}=1,$ with $A_{\mathbf{k}%
}=JM_{0}k^{2}+KM_{0}+2\pi M_{0}\left(  k_{x}^{2}+k_{y}^{2}\right)  /k^{2},$
$B_{\mathbf{k}}=2\pi M_{0}\left(  k_{x}+ik_{y}\right)  ^{2}/k^{2}\ $and
$\mathcal{H}=\mathcal{H}_{0}+\sum_{\mathbf{k}}\hslash\omega_{\mathbf{k}%
}a_{\mathbf{k}}^{+}a_{\mathbf{k}}$ with $\omega_{\mathbf{k}}=\sqrt
{A_{\mathbf{k}}^{2}-\left\vert B_{\mathbf{k}}\right\vert ^{2}}.$ The total
angular momentum Eq. \eqref{TA} reads
\begin{equation}
J_{z}-J_{z0}=-\hslash\sum_{\mathbf{k}}a_{\mathbf{k}}^{+}a_{\mathbf{k}%
}-i\hslash\sum_{\mathbf{k}}a_{\mathbf{k}}^{+}\left(  \mathbf{k}\times
\nabla_{\mathbf{k}}\right)  a_{\mathbf{k}}, \label{TAQ}%
\end{equation}
where $\nabla_{\mathbf{k}}$ is the gradient in $\mathbf{k}$\textbf{-}space.
The first/second terms on the right-hand side are the spin/orbital angular
momenta of the magnon excitations. By transforming from Cartesian to
cylindrical coordinates $\rho,$ $k_{z},$ and $n,$ and by $a_{\mathbf{k}}%
=\sum_{n}a_{\rho,k_{z},n}e^{in\phi},$ where $\rho=\sqrt{k_{x}^{2}+k_{y}^{2}}$
and $n$ is the azimuthal quantum number [the eigenvalue of the operator
$i\left(  \mathbf{k}\times\nabla_{\mathbf{k}}\right)  _{z}=i\frac{\partial
}{\partial\phi}$], $J_{z}-J_{z0}=\hslash\sum_{n}\left(  n-1\right)
a_{\rho,k_{z},n}^{+}a_{\rho,k_{z},n}$. The total angular momentum therefore
depends on the magnetization profile in the transverse plane. Different $n$
correspond to different wave-front shapes of the helical (vortex) spin wave
modes. The dipole-exchange spin waves in cylindrical ferromagnetic nanowires
display a rich wave pattern in the cross-section of the nanowire \cite{Mills}.
Vortex modes with high orbital angular momentum have been achieved in photonic
\cite{Cai} and electronic \cite{McMorran} wave guides using spiral
phase-plates, computer-generated holograms, etc. It should be very interesting
to generate helical spin wave modes and realize the conversion of angular
momentum from orbit to spin experimentally. Spin waves with high orbital
angular momenta would be efficient drivers of domain wall motion in axially
symmetric nanocylinders that have been successfully fabricated and imaged
recently \cite{Biziere}. The natural magnon mode in a cylindrical domain
carries an orbital angular momentum $n\hslash.$ Spin waves propagating through
a domain wall accumulate phase shifts \cite{Yan} corresponding to an orbital
angular momentum $n^{\prime}\hslash$ of transmitted waves ($n\neq n^{\prime}$
for complex wall structures in the presence of dipole-dipole interactions).
The transfer of orbital angular momentum\ is enhanced for large $\left\vert
n-n^{\prime}\right\vert ,$ leading to efficient domain wall motion.

To conclude, we formulate the conservation laws of linear and angular momenta
in ferromagnetic textures in the presence of magnetic dipole-dipole
interactions based on Noether's theorem. We derive a well-defined linear
momentum for insulating ferromagnets without involving incoherent fermionic
excitations, thereby resolving a paradox raised by Volovik \cite{Volovik}.
Mathematics helps to correct misguided physical intuition that naively
associates linear momentum to domain wall translational motion. Instead, we
show that linear momentum transfer of spin waves reflected at magnetic domain
walls induces an effective field and steady rotation of the domain wall plane
rather than translation. Only in the presence of dissipation this leads to
domain wall propagation. Besides the usual spin angular momentum, we identity
an orbital angular momentum for spin waves that is linked to the shape of
their wave fronts. We expect to stimulate experiments on the preparation and
manipulation of spin waves thereby opening a new research direction in the
field of magnonics.

This work is supported by the FOM foundation, DFG Priority Program 1538
SpinCat, EG-STREP MACALO, Marie-Curie ITN Spinicur and Grand-in-Aid for
Scientific Research A (Kakenhi) 25247056.


\begin{thebibliography}{99}
\bibitem {Prepare}Detailed calculations will be presented elsewhere.
\end{thebibliography}
\end{document}